\documentclass[preprint,showpacs,preprintnumbers,amsmath,amssymb]{revtex4}
\usepackage{epsfig}
\usepackage{graphicx}% Include figure files
\usepackage{dcolumn}% Align table columns on decimal point
\usepackage{bm}% bold math

\newcommand{\ord}{{\cal O}}
\def\beq{\begin{equation}}
\def\eeq#1{\label{#1}\end{equation}}
\def\eeqn{\end{equation}}
\newcommand\iden{\leavevmode\hbox{\small1\normalsize\kern-.33em1}}

\newcommand{\sq}{\sqrt{2}}
\newcommand{\bea} {\begin{eqnarray}}
\newcommand{\eea} {\end{eqnarray}}

\newcommand{\Lg}{{\mathcal L}}
\newcommand{\rd}{\partial}

\newcommand{\Gm}{\Gamma}
\newcommand{\sbt}{s_{\beta}}
\newcommand{\cbt}{c_{\beta}}
\newcommand{\tbt}{t_{\beta}}
\newcommand{\sbcb}{s_{\beta} c_\beta}
\let\jnfont=\rm
\def\NPB#1,{{\jnfont Nucl.\ Phys.\ B }{\bf #1},}
\def\PLB#1,{{\jnfont Phys.\ Lett.\ B }{\bf #1},}
\def\EPJC#1,{{\jnfont Eur.\ Phys.\ Jour.\ C }{\bf #1},}
\def\PRD#1,{{\jnfont Phys.\ Rev.\ D }{\bf #1},}
\def\PRL#1,{{\jnfont Phys.\ Rev.\ Lett.\ }{\bf #1},}
\def\MPLA#1,{{\jnfont Mod.\ Phys.\ Lett.\ A }{\bf #1},}
\def\JPG#1,{{\jnfont J.\ Phys.\ G }{\bf #1},}
\def\CTP#1,{{\jnfont Commun.\ Theor.\ Phys.\ }{\bf #1},}
\def\JHEP#1,{{\jnfont JHEP \ }{\bf #1},}
\def\NPPS#1,{{\jnfont Nucl.\ Phys.\ Proc.\ Suppl.\ }{\bf #1},}
\def\CPC#1,{{\jnfont Computl.\ Phys.\ Commun.\ }{\bf #1},}
\def\APPB#1,{{\jnfont Acta\ Phys.\ Polon.\ B }{\bf #1},}

\def\EPL#1,{{\jnfont Europhys.\ Lett. }{\bf #1},}

\begin{document}

\title{\ \\[10mm] Pseudoscalar boson and SM-like Higgs boson productions at LHC \\ in simplest little Higgs model}

\author{Lei Wang, Xiao-Fang Han}

\affiliation{ Department of Physics, Yantai University, Yantai
264005, PR China}

\begin{abstract}
In the framework of the simplest little Higgs model (SLHM), we
perform a comprehensive study for the pair productions of the
pseudoscalar boson $\eta$ and SM-like Higgs boson $h$ at LHC, namely
$gg(b\bar{b})\to \eta\eta$, $gg(q\bar{q})\to \eta h$ and
$gg(b\bar{b})\to hh$. These production processes provide a way to
probe the couplings between Higgs bosons. We find that the cross
section of $gg\to \eta\eta$ always dominates over that of
$b\bar{b}\to\eta\eta$. When the Higgs boson $h$ which mediates these
two processes is on-shell, their cross sections can reach several
thousand $fb$ and several hundred $fb$, respectively. When the
intermediate state $h$ is off-shell, those two cross sections are
reduced by two orders of magnitude, respectively. The cross sections
of $gg\to \eta h$ and $q\bar{q}\to \eta h$ are about in the same
order of magnitude, which can reach $\ord{(10^2fb)}$  for a light
$\eta$ boson. Besides, compared with the SM prediction, the cross
section of a pair of SM-like Higgs bosons production at LHC can be
enhanced sizably. Finally, we briefly discuss the observable
signatures of $\eta\eta$, $\eta h$ and $hh$ at the LHC,
respectively.

\end{abstract}


\maketitle

\section{Introduction}
Little Higgs theory \cite{LH} has been proposed as an interesting
solution to the hierarchy problem. So far various realizations of
the little Higgs symmetry structure have been proposed
\cite{otherlh,lht,lst,sst}, which can be categorized generally into
two classes \cite{smoking}. One class use the product group,
represented by the littlest Higgs model \cite{lst}, in which the SM
$SU(2)_L$ gauge group is from the diagonal breaking of two (or more)
gauge groups. The other class use the simple group, represented by
the simplest little Higgs model (SLHM) \cite{sst}, in which a single
larger gauge group is broken down to the SM $SU(2)_L$.

Since these little Higgs models mainly alter the properties of the
Higgs boson, hints of these models may be unraveled from various
Higgs boson processes. The phenomenology of Higgs boson in these
little Higgs models has been widely studied
\cite{hrrhan,higgslh,slhhdec,slhv,etaprodecay,wketa,slhhh,lhhh}. In
addition to the SM-like Higgs boson $h$, the SLHM predicts a
pseudoscalar boson $\eta$, whose mass can be as low as
$\ord{(10~\mathrm{GeV})}$. The constraint from the non-observation
in the decay $\Upsilon\to \gamma \eta$ excludes $\eta$ with mass
below 5-7 GeV \cite{5-7GeV}. In this paper, we will focus on the
pair productions of neutral Higgs bosons at LHC in the SLHM, namely
$gg(b\bar{b})\to \eta\eta$, $gg(q\bar{q})\to \eta h$ and
$gg(b\bar{b})\to hh$. These production processes at LHC are very
important because they will provide a way to probe the couplings of
$h\eta\eta$ and $hhh$, and shed light on the Higgs potential.
Further, Higg-pair production at LHC may be sensitive to new
physics, and it has been studied in many new physics model, such as
little Higgs models \cite{slhhh,lhhh}, supersymmetric models
\cite{susy}, models of universal extra dimensions \cite{UED} and
left-right twin Higgs model \cite{lrth}. Note that $gg(b\bar{b})\to
hh$ process has been studied in \cite{slhhh}. In order to perform a
comprehensive study for the pair productions of neutral Higgs bosons
at LHC, we reconsider them here.

This work is organized as follows. In Sec. II we recapitulate the
SLHM. In Sec. III we study $gg(b\bar{b})\to \eta\eta$,
$gg(q\bar{q})\to \eta h$ and $gg(b\bar{b})\to hh$ production
processes at LHC, respectively. Finally, we give our conclusion in
Sec. IV.

\section{Simplest little Higgs model}
The SLHM is based on $[SU(3) \times U(1)_X]^2$ global symmetry. The
gauge symmetry $SU(3) \times U(1)_X$ is broken down to the SM
electroweak gauge group by two copies of scalar fields $\Phi_1$ and
$\Phi_2$, which are triplets under the $SU(3)$ with aligned VEVs
$f_1$ and $f_2$. The uneaten five pseudo-Goldstone bosons can be
parameterized as

\beq
\Phi_{1}= e^{ i\; t_\beta \Theta } \left(\begin{array}{c} 0 \\
0 \\ f_1
\end{array}\right)\;,\;\;\;\;
\Phi_{2}= e^{- \; \frac{i}{t_\beta} \Theta} \left(\begin{array}{c} 0 \\  0 \\
f_2
\end{array}\right)\;,
\label{paramet}
\end{equation}
where
\begin{equation}
   \Theta = \frac{1}{f} \left[
        \left( \begin{array}{cc}
        \begin{array}{cc} 0 & 0 \\ 0 & 0 \end{array}
            & H \\
        H^{\dagger} & 0 \end{array} \right)
        + \frac{\eta}{\sqrt{2}}
        \left( \begin{array}{ccr}
        1 & 0 & 0 \\
        0 & 1 & 0 \\
        0 & 0 & 1 \end{array} \right) \right],
\end{equation}
$f=\sqrt{f_1^2+f_2^2}$ and $t_\beta\equiv \tan\beta= f_2 / f_1$.
Under the $SU(2)_L$ SM gauge group, $\eta$ is a real scalar, while
$H$ transforms as a doublet and can be identified as the SM Higgs
doublet. The kinetic term in the non-linear sigma model is \beq
\label{eq:Lg:gauge0} \Lg_\Phi = \sum_{j=1,2}\left| \left(\rd_\mu + i
g A^a_\mu T^a - i \frac{g_x}{3} B^x_\mu \right) \Phi_j \right|^2,
\end{equation}
where $g_x =g\tan \theta_W/ \sqrt{1-\tan^2 \theta_W/3}$ with
$\theta_W$ being the electroweak mixing angle. As $\Phi_1$ and
$\Phi_2$ develop their VEVs, the new heavy gauge bosons $Z'$, $Y^0$,
and $X^{\pm}$ get their masses proportional to $f$: \beq
m^2_{Z'}=g^2f^2\frac{2}{3-\tan^2\theta_W},~~~m^2_{X^{\pm}} =
m^2_{Y^{0}} = \frac{g^2}{2}f^2.
\end{equation}

The gauged $SU(3)$ symmetry promotes the SM fermion doublets into
$SU(3)$ triplets. There are two possible gauge charge assignments
for the fermions: the 'universal' embedding and the 'anomaly-free'
embedding.  The first choice is not favored by the
electroweak precision data \cite{sst}, so we focus on the second
way of embedding. The quark Yukawa interactions for the third generation
and the first two generations can be written respectively as
\bea
{\cal L}_3 &=& i \lambda_1^t t_1^c \Phi_1^{\dagger} Q_3
  + i \lambda_2^t t_2^c \Phi_2^{\dagger} Q_3
  + i  \frac{\lambda_d^m}{\Lambda}  d_m^c \epsilon_{ijk}
      \Phi_1^i \Phi_2^j Q_3^k + h.c., \label{simtopyukawa}\\
{\cal L}_{1,2} &=&  i \lambda_1^{d_n} d_{1n}^c Q_n^{T} \Phi_1
  + i \lambda_2^{d_n} d_{2n}^c Q^{T}_n \Phi_2
  + i \frac{\lambda_{u}^{mn}}{\Lambda} u_m^c \epsilon_{ijk} \Phi_1^{*i}
    \Phi_2^{*j} Q_n^k + h.c.,\label{simucyukawa}
\eea where $n=1,2$ are the first two generation indices;
$i,j,k=1,2,3$; $Q_3=\{ t_L, b_L, i T_L\}$ and $Q_n = \{ d_{nL}, -
u_{nL}, i D_{nL}\}$; $d_m^c$ runs over $(d^c, s^c, b^c, D^c, S^c)$;
$d^c_{1n}$ and $d^c_{2n}$ are linear combinations of $d^c$ and $D^c$
for $n=1$ and of $s^c$ and $S^c$ for $n=2$; $u^c_m$ runs over $(u^c,
c^c, t^c, T^c)$. For simplicity, we assume the quark flavor mixing
are small and neglect the mixing effects. Eqs. (\ref{simtopyukawa})
and (\ref{simucyukawa}) contain the Higgs boson interactions and the
mass terms for the three generations of quarks: \bea
\label{tTmixing} {\cal L}_t &\simeq&-f \lambda_2^t \left[
x_\lambda^t c_\beta  t_1^c(-s_1t_L
   +c_1T_L)G_1(\eta)+s_\beta t_2^c (s_2 t_L+ c_2 T_L) G_2(\eta) \right]+h.c.,\,\\
   \label{dDmixing}
{\cal L}_{d_n} &\simeq&-f \lambda_2^{d_n} \left[ x_\lambda^{d_n} c_\beta d_1^c
  (s_1 d_{nL}+c_1 D_{nL})G^*_1(\eta)+s_\beta d_2^c (-s_2 d_{nL}+c_2 D_{nL})G^*_2(\eta)\right]+h.c.,\,\\
{\cal L}_{b} &\simeq&-\frac{\lambda_b}{\Lambda}f^2 s_\beta c_\beta
s_3 b^c b_{_L}G_3(\eta)
          +h.c.,\,\\
{\cal L}_{q} &\simeq&-\frac{\lambda_q}{\Lambda}f^2 s_\beta c_\beta
s_3 q^c q_{_L}G^*_3(\eta)+h.c. \ (q=u,c),\, \eea where
 \bea x_\lambda^t&\equiv & {\lambda_1^t \over \lambda_2^t},\
\ x_\lambda^{d_n}\equiv {\lambda_1^{d_n} \over \lambda_2^{d_n}},\ \
s_{\beta}\equiv\frac{f_2}{\sqrt{f^2_1+f^2_2}},\ \
c_{\beta}\equiv\frac{f_1}{\sqrt{f^2_1+f^2_2}},\ \ \nonumber\\
s_1&\equiv & \sin {t_\beta (h+v)\over \sqrt{2}f},\ \ s_2\equiv
\sin{(h+v) \over \sqrt{2}t_\beta f},\ \ s_3\equiv
\sin{(h+v)(t_\beta^2+1)\over \sqrt{2}t_\beta f},\ \ \nonumber\\
G_1(\eta)&\equiv &
1-i\frac{t_{\beta}}{\sqrt{2}f}\eta-\frac{t_{\beta}^2}{4f^2}\eta^2,\
\
G_2(\eta)\equiv 1+i\frac{1}{\sqrt{2}t_{\beta}f}\eta-\frac{1}{4t_{\beta}^2f^2}\eta^2,\ \ \nonumber\\
G_3(\eta)&\equiv &
1+i\frac{1}{\sqrt{2}f}(t_\beta-\frac{1}{t_\beta})\eta
-\frac{1}{4f^2}(t_\beta-\frac{1}{t_\beta})^2\eta^2, \eea with $h$
and $v$ being the SM-like Higgs boson field and its VEV,
respectively. The mass eigenstates are obtained by mixing the
corresponding interaction eigenstates, e.g., the mass eigenstates
$(t_{mL}, T_{mL})$ and $(t_m^c, T_m^c)$ are  respectively the
mixtures of $(t_{L}, T_{L})$ and $(t^c, T^c)$. The diagonalization
of the mass matrix in Eqs. (\ref{tTmixing}) and (\ref{dDmixing}) is
performed numerically in our analysis, and the relevant couplings of
$h$ and $\eta$ bosons can also be obtained without resort to any
expansion of $v/f$. Hereafter we denote the mass eigenstates without
the subscript '$m$' for simplicity.

The Yukawa and gauge interactions break the global symmetry and then
provide a potential for the Higgs boson. However, the
Coleman-Weinberg potential alone is not sufficient since the
generated $h$ mass is too heavy and the new pseudoscalar $\eta$ is
massless. Therefore, one can introduce a tree-level $\mu$ term which
can partially cancel the $h$ mass \cite{sst,slhv}: \beq -\mu^2
(\Phi^\dagger_1 \Phi_2 + h.c.) = - 2 \mu^2 f^2 \sbt\cbt \cos\left(
\frac{\eta}{\sq \sbt\cbt f} \right)
 \cos \left(
 \frac{\sqrt{H^\dagger H}}{f \cbt\sbt}
\right).
\end{equation}
The Higgs potential becomes \beq \label{eq:VCW} V = - m^2 H^\dagger
H + \lambda (H^\dagger H)^2
 - \frac{1}{2} m_\eta^2 \eta^2 +\lambda' H^\dagger H \eta^2 + \cdots,
\end{equation}
where
\beq \label{eq:msq:lambda}
m^2 = m_0^2 - \frac{\mu^2}{\sbcb}, \quad
\lambda =\lambda_0 - \frac{\mu^2}{12\sbt^3 \cbt^3f^2}, \quad
\lambda' = - \frac{\mu^2}{4 f^2 \sbt^3 \cbt^3},
\end{equation}
with $m_0$ and $\lambda_0$ being respectively the one-loop
contributions to the $h$ mass and the quartic couplings from the
contributions of fermion loops and gauge boson loops \cite{sst}. The
Higgs VEV and the masses of $h$ and $\eta$ are given by \beq
\label{eq:vsq:mH:meta} v^2 = \frac{ m^2}{\lambda} , \quad m_h^2 = 2
m^2 , \quad m_\eta^2 = \frac{\mu^2}{\sbcb} \cos\left(
\frac{v}{\sqrt{2} f \sbcb} \right).
\end{equation}
The Coleman-Weinberg potential involves the following parameters:
\beq \label{para}
f,~ x_\lambda^t,~ t_\beta,~\mu,~m_\eta,~m_h,v.
\end{equation}
Due to the modification of the observed $W$ gauge boson mass, $v$ is
defined as \cite{slhv} \beq \label{eq:v} v \simeq v_0 \left[ 1+
\frac{v_0^2}{12 f^2}\frac{\tbt^4-\tbt^2+1}{\tbt^2} -
\frac{v_0^4}{180 f^4}\frac{\tbt^8-\tbt^6+\tbt^4-\tbt^2+1}{\tbt^4}
\right],
\end{equation}
where $v_0=246$ GeV is the SM Higgs VEV. Assuming that there are no
large direct contributions to the potential from physics at the
cutoff, we can determine other parameters in Eq. (\ref{para}) from
$f$, $t_\beta$ and $m_{\eta}$ ($m_h$) with the definition of $v$ in
Eq. (\ref{eq:v}).

\section{$\eta \eta$, $\eta h$ and $hh$ productions at LHC}
\begin{figure}[tb]
\begin{center}
 \epsfig{file=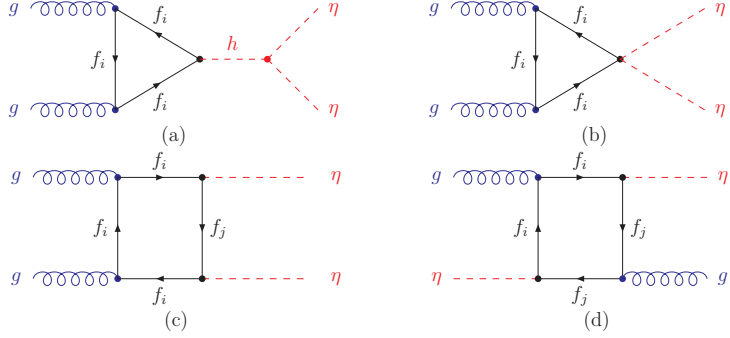,width=10cm}
\end{center}
\vspace{-1.0cm} \caption{Feynman diagrams for double $\eta$
production via gluon-gluon fusion in the SLHM. Here $i,j=1,2$ with
$(f_1,f_2)$ denoting $(t, T)$ or $(d, D)$ or $(s, S)$. The diagrams
by exchanging the two gluons or exchanging the two $\eta$ bosons in
(c,d) are not shown here.} \label{fmgg}
\end{figure}

\begin{figure}[tb]
\begin{center}
 \epsfig{file=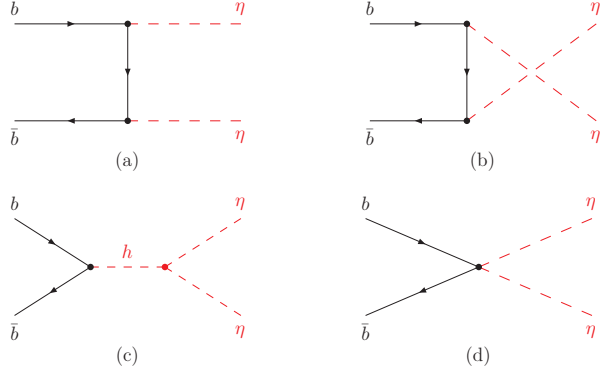,width=8cm}
\end{center}
\vspace{-1.0cm} \caption{Feynman diagrams for double $\eta$
production via $b\bar{b}$ annihilation in the SLHM. } \label{fmbb}
\end{figure}

At the LHC the double $\eta$ production can proceed through
gluon-gluon fusion and $b\bar{b}$ annihilation, as shown in Fig.
\ref{fmgg} and Fig. \ref{fmbb}, respectively. Since their Yukawa
couplings are very small, we do not consider the contributions of
$q\bar{q}~(q=u,c,d,s)$ annihilation processes. For the gluon-gluon
fusion process, there are two types of Feynman diagrams. One is the
triangle diagrams where an off-shell (or on-shell) $h$ boson,
produced from gluon-gluon fusion through the heavy quark loops,
decays into a pair of $\eta$ bosons.
 The other is the box diagrams where the double $\eta$
bosons are produced through quark boxes. Due to the large Yukawa
couplings of the heavy quarks and the large parton distribution
function of gluon at the LHC, the contributions of the gluon-gluon
fusion process can dominate over those of $b\bar{b}$ annihilation
process. The Feynman diagrams of $gg(b\bar{b})\to hh$ process can be
obtained from those of  $gg(b\bar{b})\to \eta\eta$ by replacing the
final state two $\eta$ bosons with two $h$ bosons.

\begin{figure}[tb]
\begin{center}
 \epsfig{file=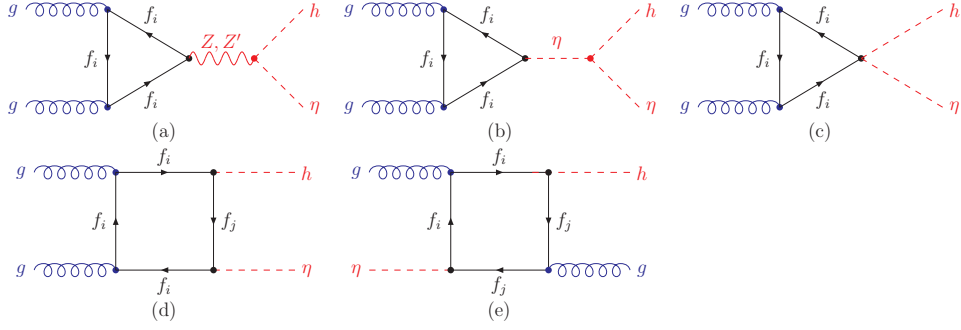,width=13cm}
\end{center}
\vspace{-1.0cm} \caption{Feynman diagrams for $\eta h$ associated
production via gluon-gluon fusion in the SLHM. For (b-e), $i,j=1,2$
with $(f_1,f_2)$ denoting $(t, T)$ or $(d, D)$ or $(s, S)$ ; For
(a), $f_i$ denotes the SM quarks and new quarks $T$, $D$ and $S$.
The diagrams by exchanging the two gluons or exchanging the $\eta$
and $h$ bosons in (d,e) are not shown here.} \label{fmggyh}
\end{figure}

\begin{figure}[tb]
\begin{center}
 \epsfig{file=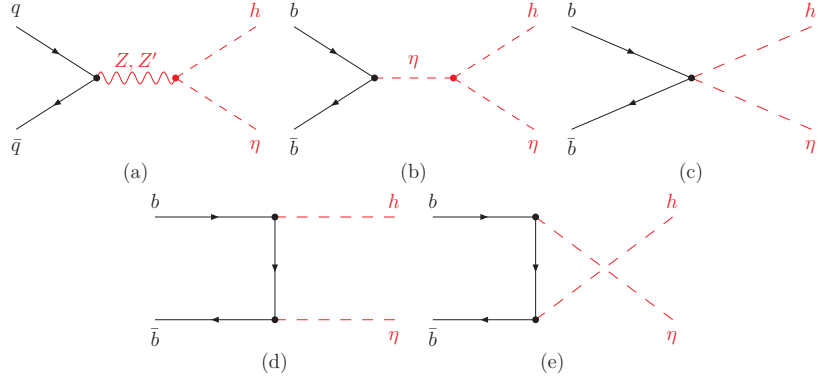,width=11cm}
\end{center}
\vspace{-1.0cm} \caption{Feynman diagrams for $\eta h$ associated
production via $b\bar{b}$ annihilation in the SLHM. The parton $q$
in (a) denotes $u,~c,~d,~s,~b$.} \label{fmbbyh}
\end{figure}

The $\eta h$ associated production at LHC can proceed through
gluon-gluon fusion and $q\bar{q}$ annihilation, as shown in Fig.
\ref{fmggyh} and Fig. \ref{fmbbyh}. In addition to the diagrams
which are similar to those of $\eta\eta$ production, there are some
Feynman diagrams where $Z$ and $Z'$, produced from gluon-gluon
fusion through the quark loops and from $q\bar{q}$ annihilation,
decay into $\eta h$ as shown in Fig. \ref{fmggyh}(a) and Fig.
\ref{fmbbyh}(a), respectively. Although the new neutral gauge boson
$Y^0$ can also contribute to the processes $gg\to \eta h$ and
$q\bar{q}\to \eta h$, its gauge coupling is suppressed by $v/f$ and
$1/t_\beta$ \cite{smoking,etaprodecay}. Therefore, we neglect the
contributions of $Y^0$.

%%%%%%%%%%%%%%%%%%%%%
%%%%%%%%%%%%%%%%%%%%
%%%%%%%%%%%%%%%%%%%%%
%%%%%%%%%%%%%%%%%%%%

%%%%%%%%%%%%%%%%%%%%%
%%%%%%%%%%%%%%%%%%%%
%%%%%%%%%%%%%%%%%%%%%
%%%%%%%%%%%%%%%%%%%%

The calculations of the loop diagrams in Fig. \ref{fmgg} and Fig.
\ref{fmggyh} are straightforward. Each loop diagram is composed of
some scalar loop functions \cite{Hooft} which are calculated by
using LoopTools \cite{Hahn}. The amplitudes of triangle diagrams are
as follows: \bea M_{\ref{fmgg}(\rm a)}+M_{\ref{fmgg}(\rm
b)}&=&\frac{\delta^{ab}g_s^2}{4\pi^2}\epsilon^a_1(k_1)\cdot\epsilon^b_2(k_2)\sum_f
m_f (1+2m_f^2C_0-k_1\cdot k_2C_0)\left(\frac{g_{h\bar{f}f}\cdot
g_{h\eta\eta}}{\hat{s}-m_h^2+im_h\Gamma_h}+g_{\eta\eta\bar{f}f}\right),\nonumber \\
M_{\ref{fmggyh}(\rm b)}+M_{\ref{fmggyh}(\rm
c)}&=&\frac{\delta^{ab}g_s^2}{4\pi^2}\epsilon^{\mu}_1(k_1)\epsilon^{\nu}_2(k_2)
k^{\beta}_1k^{\alpha}_2 \varepsilon_{\mu\alpha\nu\beta}  \sum_f m_f
C_0\left(\frac{g_{\eta\bar{f}f}\cdot
g_{h\eta\eta}}{\hat{s}-m_{\eta}^2}+g_{h\eta\bar{f}f}\right),
 \eea
where $\hat{s}=k^2=(k_1+k_2)^2$. $M_{x(y)}$ denotes the amplitude of
Fig. $x(y)$ with $x=1,~3$ and $y$=a,~b,~c. $C_0\equiv
C_0(0,0,\hat{s},m_f^2,m_f^2,m_f^2)$ is the 3-point Feynman integrals
scalar function, and $a$ and $b$ denote the color factor of gluons.
\bea M_{\ref{fmggyh}(\rm a)}&=&\frac{-g_{\alpha\beta}+k_\alpha
k_\beta/m_Z^2}{\hat{s}-m_Z^2}g_{_{Z\eta h}}(p_h-p_\eta)^\beta
\epsilon_{1\mu}(k_1)\epsilon_{2\nu}(k_2)\sum_f
F^{\alpha\mu\nu}(m_f,g_a^{Z\bar{f}f})\\
&+&\frac{-g_{\alpha\beta}+k_\alpha
k_\beta/m_{Z'}^2}{\hat{s}-m_{Z'}^2+im_{Z'}\Gamma_{Z'}}g_{_{Z'\eta
h}}(p_h-p_\eta)^\beta \epsilon_{1\mu}(k_1)\epsilon_{2\nu}(k_2)\sum_f
F^{\alpha\mu\nu}(m_f,g_a^{Z'\bar{f}f}). \eea
$F^{\alpha\mu\nu}(m_f,g_a^{Z\bar{f}f})$ is the effective coupling of
$ggZ$ \cite{ggz,f1-f4}: \bea F^{\alpha\mu\nu}=\sum_Q
\frac{g_ag_s^2Tr[T^aT^b]}{4\pi^2}
[\varepsilon^{\mu\nu\omega\varphi}k_{1\omega}k_{2\varphi}k^{\alpha}
F_1(\hat{s})+(\varepsilon^{\alpha\mu\omega\varphi}k_2^{\nu}
-\varepsilon^{\alpha\nu\omega\varphi}k_1^{\mu})k_{1\omega}k_{2\varphi}
F_2(\hat{s}) \\ \nonumber
+(\varepsilon^{\alpha\mu\omega\varphi}k_1^{\nu}
-\varepsilon^{\alpha\nu\omega\varphi}k_2^{\mu})k_{1\omega}k_{2\varphi}
F_3(\hat{s})+\varepsilon^{\alpha\mu\nu\omega}(k_{1\omega}-k_{2\omega})
F_4(\hat{s})] \label{ggZ} \eea where $g_a$ is the coupling of axial
vector current and $F_i(\hat{s})$ ( $i=1-4$ ) are scalar functions,
\bea F_1 &=&
-\frac{1}{\hat{s}}(B_0[0,m_f^2,m_f^2]-B_0[\hat{s},m_f^2,m_f^2]+1+2C_0[0,0,\hat{s},m_f^2,m_f^2,m_f^2]m_f^2),
\nonumber\\
 -F_2 &=& F_3=\frac{2}{\hat{s}}\left[
-\frac{1}{2}(B_0[0,m_f^2,m_f^2]-B_0[\hat{s},m_f^2,m_f^2]+1
-2C_0[0,0,\hat{s},m_f^2,m_f^2,m_f^2]m_f^2)+1 \right],
\nonumber\\
F_4 &=&
-\frac{1}{2}(B_0[0,m_f^2,m_f^2]-B_0[\hat{s},m_f^2,m_f^2]+1-2C_0[0,0,\hat{s},m_f^2,m_f^2,m_f^2]m_f^2)+1,
\eea where the unity in $F_4$ is the anomaly term.

The amplitudes expressions of box diagrams are lengthy, which are
not presented here. The hadronic cross section at the LHC is
obtained by convoluting the parton cross section with the parton
distribution functions. In our calculations we use CTEQ6L
\cite{cteq} to generate the parton distributions with the
renormalization scale $\mu_R $ and the factorization scale $\mu_F$
chosen to be $\mu_R = \mu_F = 2m_{\eta}$ for $\eta\eta$ production
process ($\mu_R = \mu_F = m_{\eta}+m_h$ for $\eta h$ production;
$\mu_R = \mu_F = 2m_{h}$ for $hh$ production) and use the two-loop
running coupling constant $\alpha_{s}$ with
$\alpha_{s}(m_{Z})=0.118$. The $h$ boson mediating $gg(b\bar{b})\to
\eta\eta$ and $Z'$ mediating $gg(q\bar{q})\to \eta h$ can be
respectively on-shell for $m_{h}\geq 2m_{\eta}$ and $m_{Z'}\geq
m_{\eta}+m_h$. In order to take into account possible resonance
effects of $h$ and $Z'$, we have calculated the decay modes of $h$
and $Z'$, which are shown in appendix.

The SM input parameters relevant in our study are taken as
$m_t=173.3$ GeV \cite{10073178} and $m_{Z}=91.1876$ GeV \cite{pdg}.
The free SLHM parameters are $f, ~t_\beta,~ m_{\eta}~(m_h),
x_\lambda^{d}$ and $x_\lambda^{s}$. As shown above, the parameters
$x_\lambda^t,~\mu,~m_h~(m_\eta)$ can be determined by $f$,
$t_\beta$, $m_{\eta}$ $(m_h)$ and $v$. To satisfy the bound of LEP2,
we require that $m_h$ is larger than 114.4 GeV \cite{lep2}.
Certainly, due to the presence of the dominant decay mode $h\to
\eta\eta$ and suppression of $hZZ$ coupling \cite{slhv,slhhdec}, the
LEP2 bound on $m_h$ should be loosened to some extent. The recent
studies about $Z$ leptonic decay and $e^+e^-\to \tau^+\tau^-\gamma$
process at the $Z$ pole show that the scale $f$ should be
respectively larger than 5.6 TeV and 5.4 TeV, which does not depend
on $t_\beta$ \cite{f5.65.4}. Such large values of $f$ can suppress
the SLHM predictions sizably. However, the factor $t_{\beta}$ in the
couplings of $h$ and $\eta$ can be taken as a large value to cancel
the suppression of $f$ partially. For the perturbation to be valid,
$t_{\beta}$ cannot be too large for fixed $f$. If we require
$\ord(v_0^4/f^4)/\ord(v_0^2/f^2) < 0.1$ in the expansion of $v$,
$t_\beta$ should be below 28 for $f=5.6$ TeV. In our calculation, we
take $f=5.6$ TeV and $t_{\beta}=15,~20,~25$, respectively.

 Besides, the SLHM predicts a heavy neutrino for leptons of each generation,
and the mixing of the heavy neutrinos with the light neutrinos in
conjunction with a family mixing in the lepton sectors produces the
new lepton mixing matrix $V_{\ell}$ \cite{smoking,lpv1,lpv2}, which
can lead to lepton-flavor violating processes, such as $\mu\to
e\gamma$, $\mu\to ee\bar{e}$ and $\mu N\to e N$. The experimental
constraints from the three processes are very strong. For example,
using only two lepton generations, we need $f\gtrsim 8$ TeV or very
small mixing angles or heavy neutrinos mass splitting, i.e., $\sin
2\theta \lesssim 0.01$ or $\delta \lesssim 1\%$ with $t_\beta =1$
\cite{lpv1}. However, as in the quark sector of the SM, this
lepton-flavor violation will vanish in the limit that the mixing
matrix $V_{\ell}$ is diagonal or the masses of the heavy neutrinos
are degenerate \cite{smoking}. In this paper, we assume the two
scenarios, so that $f$ and $t_{\beta}$ are free from the
experimental constraints of the lepton-flavor violating processes.
Note that the parameters of lepton sector, i.e., mixing matrix
$V_{\ell}$ and masses of the heavy neutrinos, are not involved in
our calculations directly.

The small mass of the $d$($s$) quark requires one of the couplings
$\lambda^{d}_1$ and $\lambda^{d}_2$ ($\lambda^{s}_1$ and
$\lambda^{s}_2$) to be very small, so there is almost no mixing
between the SM down-type quarks and their heavy partners. We assume
$\lambda^{d}_1$($\lambda^{s}_1$) is small, and take
$x_\lambda^{d}=1.1\times 10^{-4}$ $(x_\lambda^s=2.1\times10^{-3})$,
which can make the masses of $D$ and $S$ be in the range of $1$ TeV
and $2$ TeV for other parameters taken in our calculations. In fact,
our results show that different choices of $x_\lambda^{d}$ and
$x_\lambda^{s}$ can not have sizable effects on the result.

%%%%%%%%%%%%%%%%%%%%%
\begin{figure}[tb]
\begin{center}
 \epsfig{file=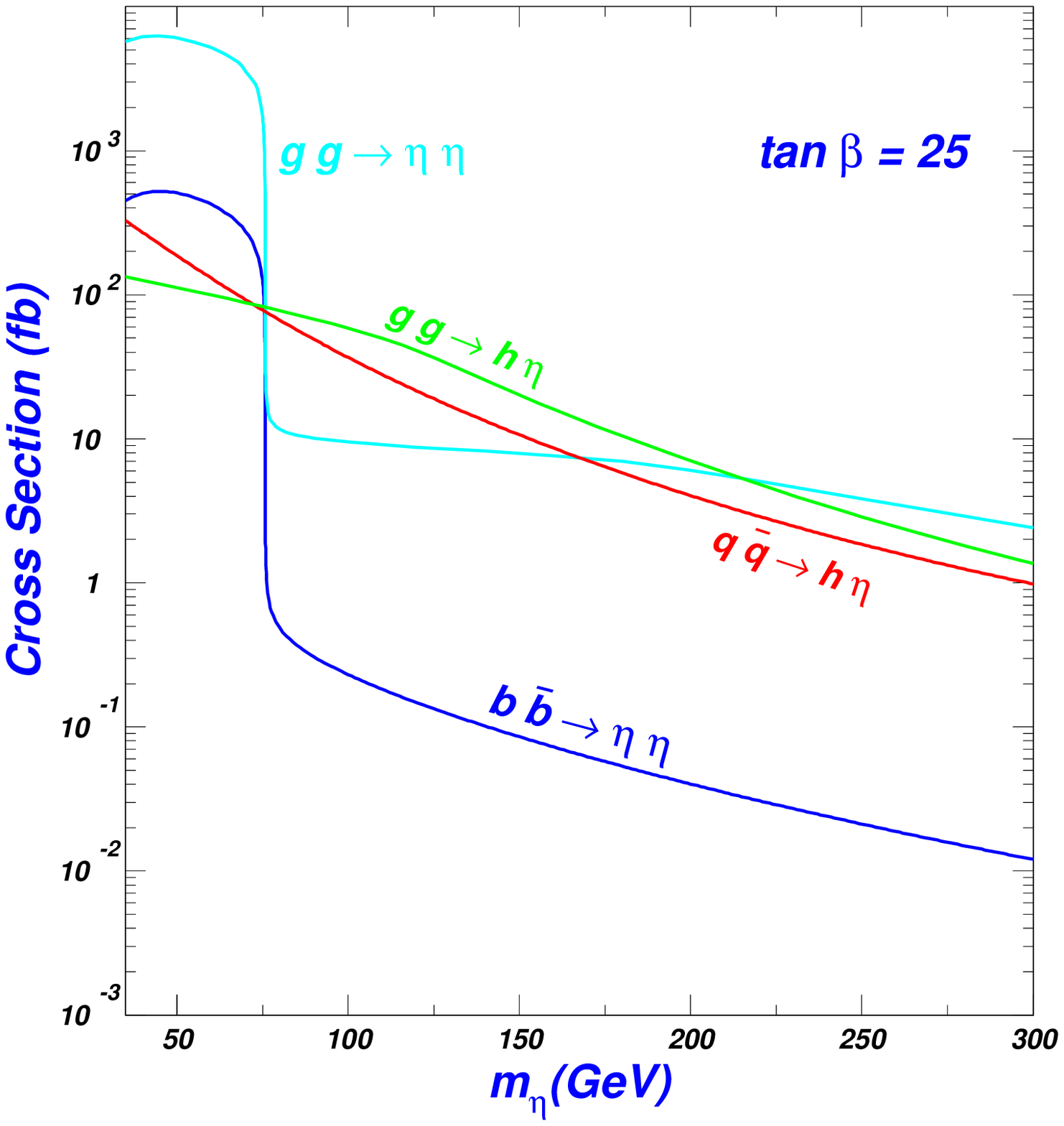,width=5.55cm}
 \epsfig{file=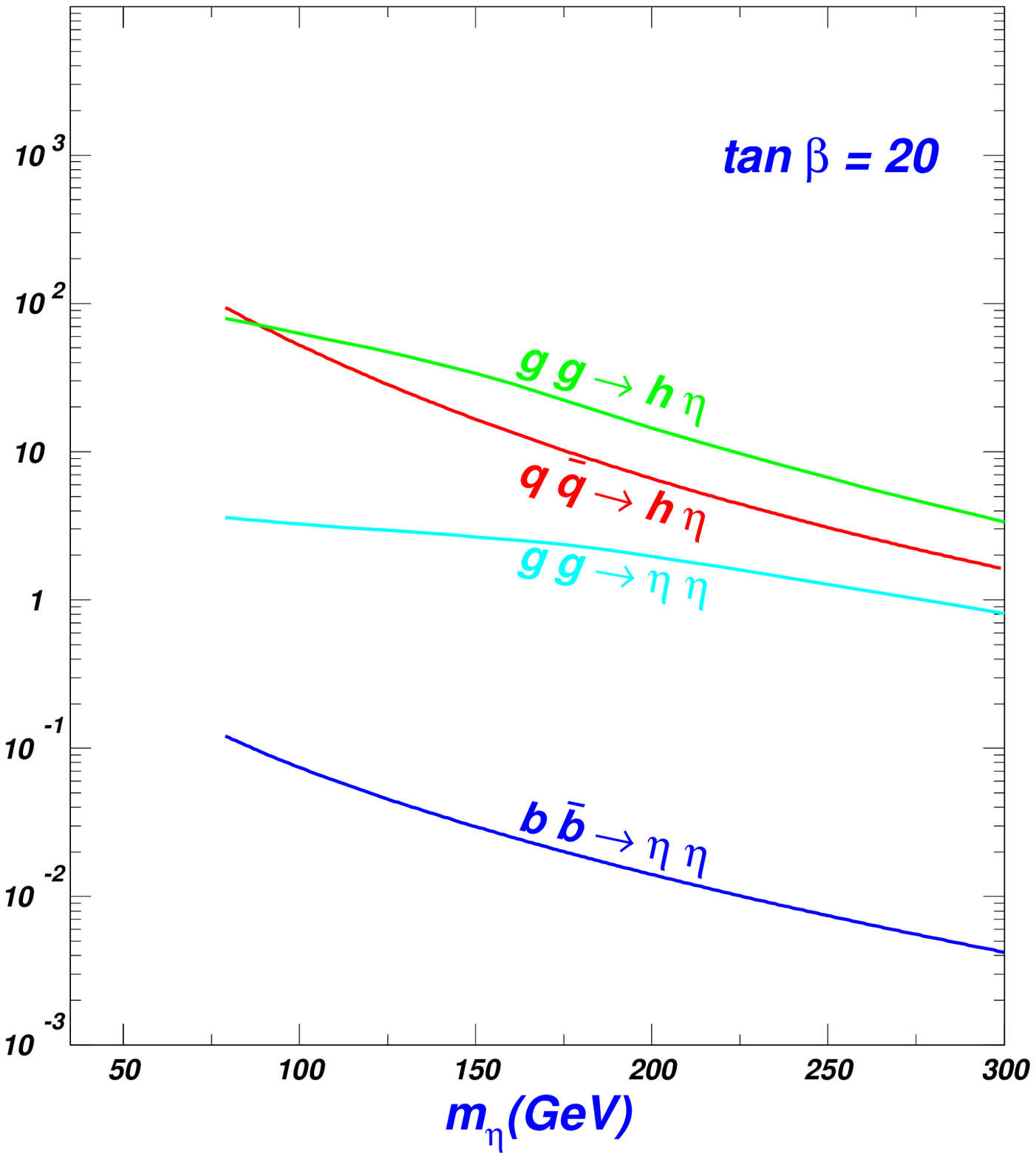,width=5.3cm}
  \epsfig{file=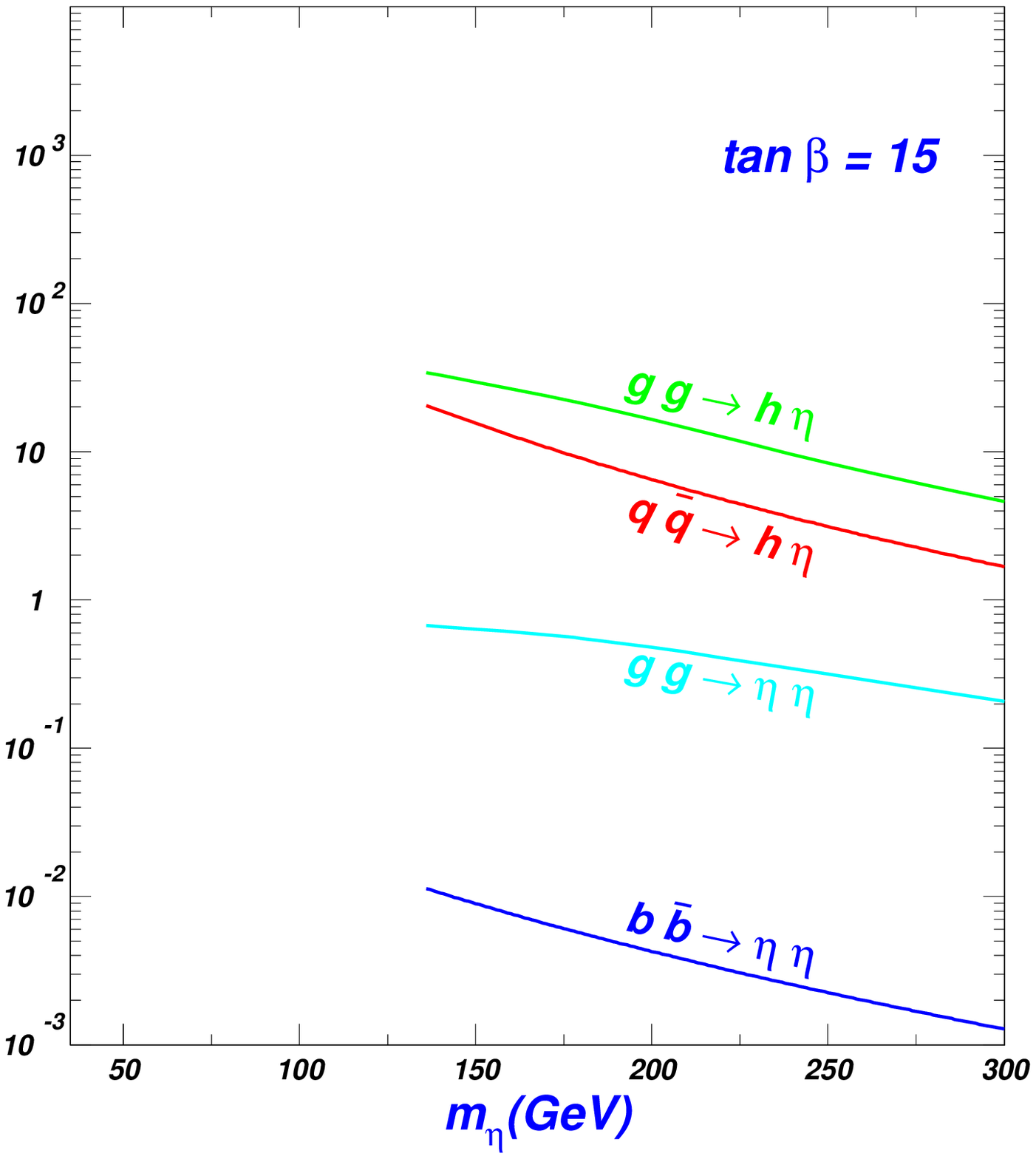,width=5.3cm}
\end{center}
\vspace{-1.0cm} \caption{For $f=5.6$ TeV, hadronic cross sections of
$gg(b\bar{b})\to \eta\eta$ and $gg(q\bar{q})\to \eta h$ at the LHC
versus the $\eta$ boson mass. The incomplete lines for
$\tan\beta=20$ and $15$ show the lower bounds of the $\eta$ mass,
respectively.} \label{cross}
\end{figure}
%%%%%%%%%%%%%%%%%%%%

In Fig. \ref{cross}, we plot the hadronic cross sections of
$gg(b\bar{b})\to \eta\eta$ and $gg(q\bar{q})\to \eta h$ at the LHC
versus the $\eta$ boson mass, respectively. Fig. \ref{cross} shows
the cross sections of these processes are all sensitive to the
$\eta$ mass, and the values decrease with increasing of the $\eta$
mass. For the double $\eta$ production, the cross section of
gluon-gluon fusion process always dominates over that of $b\bar{b}$
annihilation process. For $t_{\beta}=25$ and 35 GeV $< m_{\eta} <$
75 GeV, an on-shell $h$ boson can be produced from the gluon-gluon
fusion and $b\bar{b}$ annihilation, and decays into $\eta\eta$.
Therefore, the cross sections of the two production processes are
enhanced sizably and reach $\ord{(10^3 fb)}$ and $\ord{(10^2 fb)}$,
respectively. However, in the parameter space where the $h$ boson is
off-shell, those two cross sections can reach respectively $\ord{(10
fb)}$ and $\ord{(1 fb)}$ for $t_{\beta}=25$, and become smaller for
$t_{\beta}=20$ and $t_{\beta}=15$.

For the $\eta h$ associated production, the cross sections of
gluon-gluon fusion process and $q\bar{q}$ annihilation process are
about in the same order of magnitude, which can reach
$\ord{(10^2fb)}$ for a light $\eta$ boson. Our numerical results
show the contributions from the process mediated by $Z$ and $Z'$ as
shown in Fig. \ref{fmbbyh}(a) are dominant for $q\bar{q}\to \eta h$,
but in Fig. \ref{fmggyh}(a) are much smaller than other processes
for $gg\to \eta h$. Although the intermediate state $Z'$ can induce
the resonance effects, the contributions from the process mediated
by $Z$ can still dominate over those of $Z'$ due to the large mass
of $Z'$ and the small branching ratio of $Z'\to \eta h$.

%%%%%%%%%%%%%%%%%%%%%
\begin{figure}[tb]
\begin{center}
 \epsfig{file=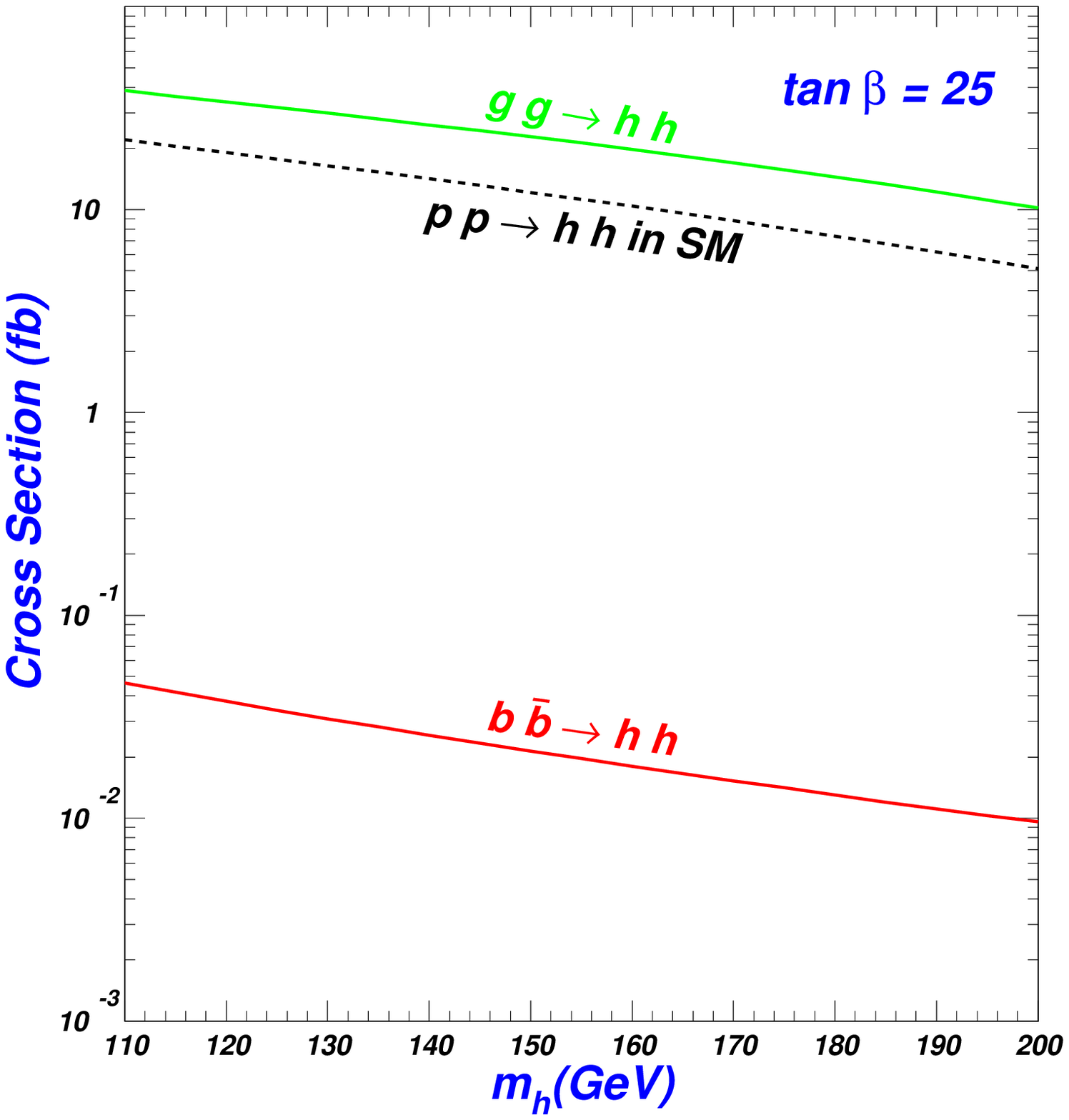,width=5.55cm}
 \epsfig{file=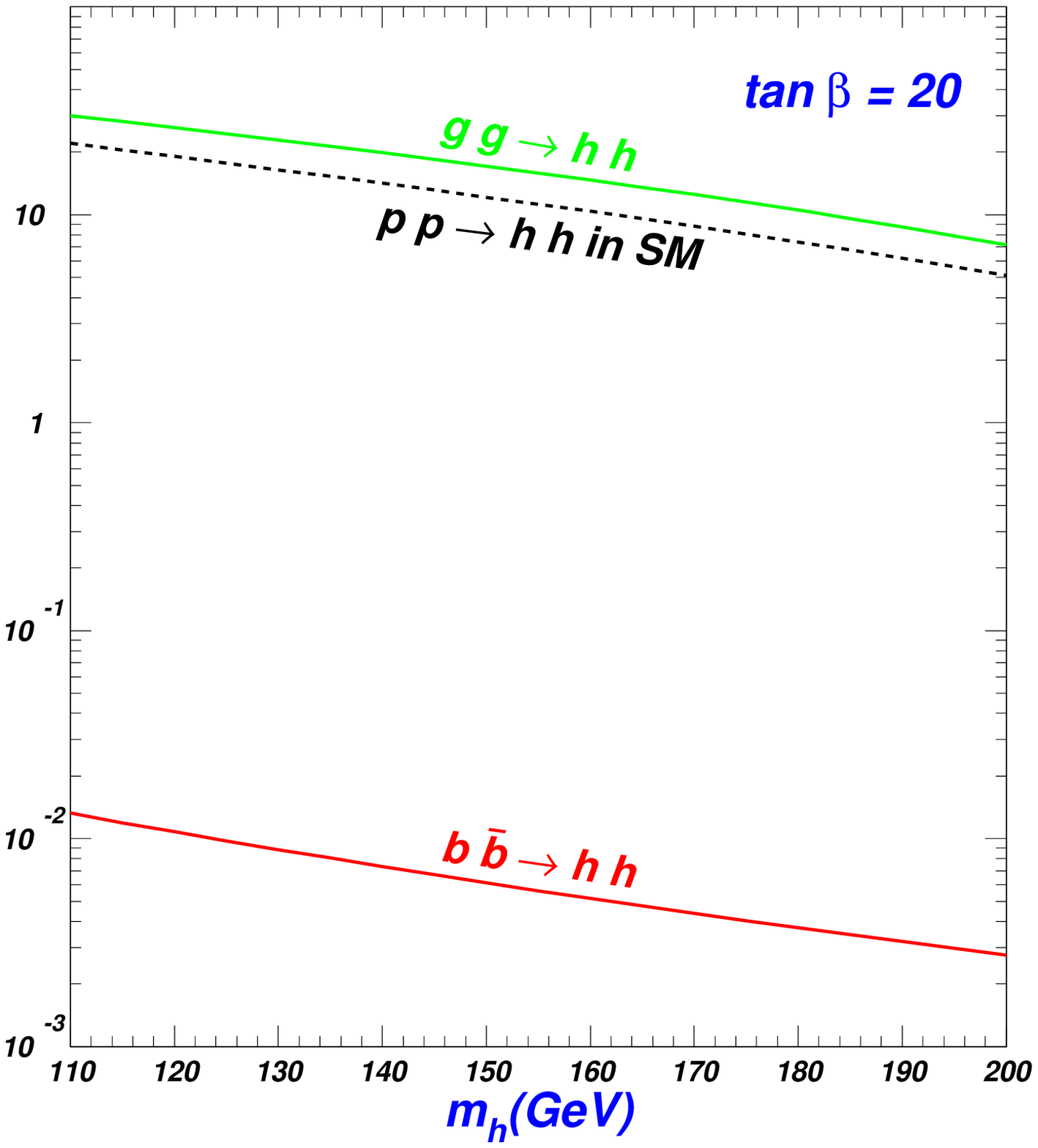,width=5.3cm}
  \epsfig{file=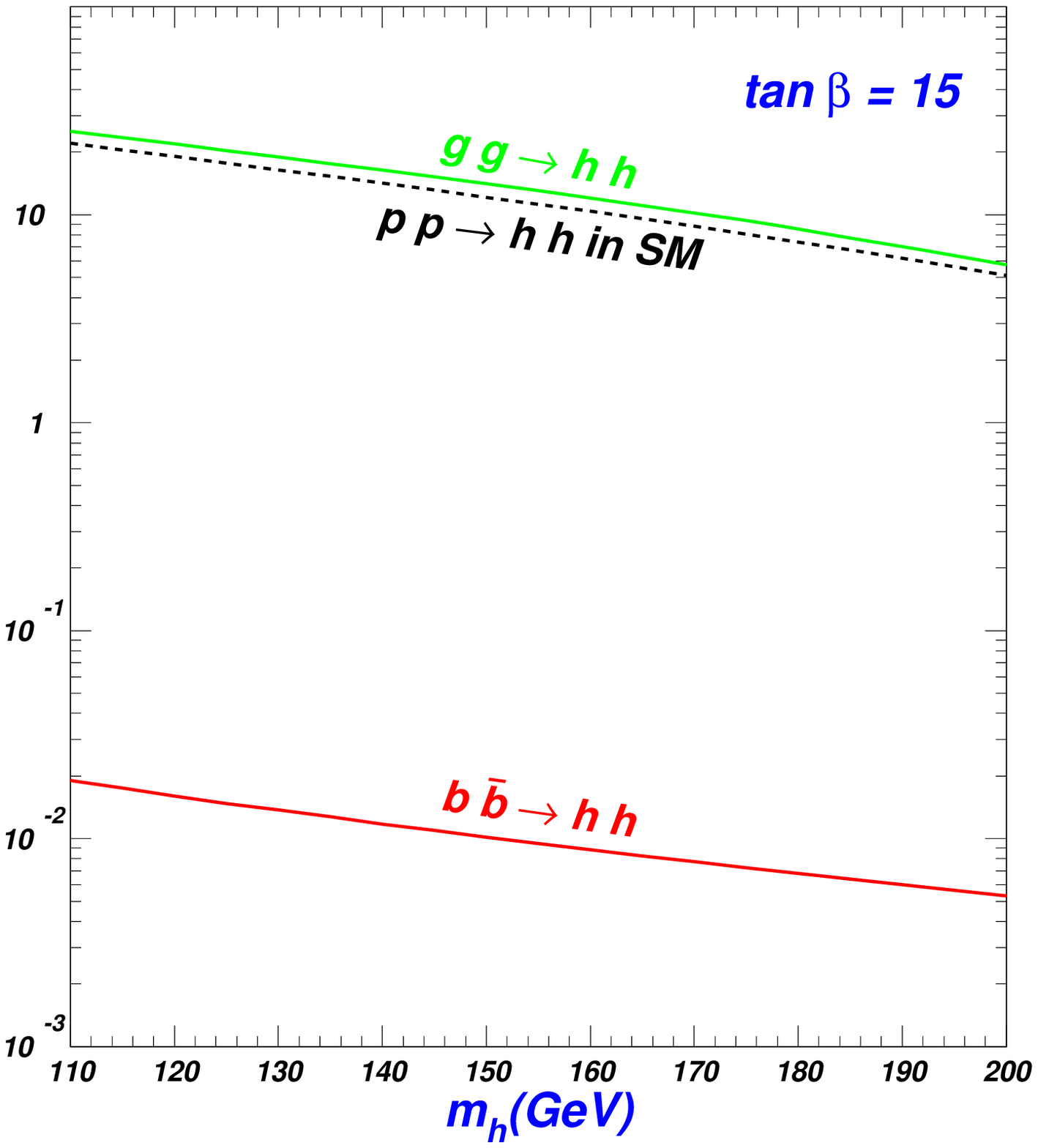,width=5.3cm}
\end{center}
\vspace{-1.0cm} \caption{For $f=5.6$ TeV, hadronic cross sections of
$gg\to hh$ and $b\bar{b}\to hh$ at the LHC versus the $h$ boson
mass. The $pp\to hh$ process in SM includes $gg\to hh$ and $b\bar{b}
\to hh$. } \label{crosshh}
\end{figure}
%%%%%%%%%%%%%%%%%%%%

In Fig. \ref{crosshh}, we plot the hadronic cross sections of $gg\to
hh$ and $b\bar{b}\to hh$ at the LHC versus the $h$ boson mass. We
find that the cross section of $b\bar{b}\to hh$ can be neglected
compared with that of $gg\to hh$, and the cross sections of the two
processes decrease with increasing of $m_h$. Compared with the SM
prediction, the cross section of $hh$ production at LHC in SLHM can
be enhanced sizably for a large $t_\beta$. For example, with $f=5.6$
TeV and $t_\beta=25$, the cross section can be approximately
enhanced by $80\%$ for $m_h = 120$ GeV.

Now we briefly discuss the observable signatures of $\eta\eta$,
$\eta h$ and $hh$ productions at LHC, respectively. For 10 GeV $<
m_\eta <$ 300 GeV, the $\eta$ boson mainly decays into $b\bar{b}$,
$\tau\bar{\tau}$ and $gg$. The branching ratio of $\eta\to
\tau\bar{\tau}$ is about $10\%$ of $\eta \to b\bar{b}$
\cite{slhv,wketa}. The huge QCD backgrounds make it essentially
impossible to discover the signatures $pp\to \eta\eta \to
b\bar{b}b\bar{b}$ and $pp\to \eta\eta \to gggg$ at LHC. If one of
the double $\eta$ bosons decays into $\tau\bar{\tau}$, and the other
decays into $b\bar{b}$, the signal to background ratio S/B can be
enhanced sizably \cite{higgsself}. Namely, $pp\to \eta\eta \to
\tau\bar{\tau}b\bar{b}$ is a promising channel to search for
$\eta\eta$ at LHC. Besides, the rare mode $pp\to \eta\eta\to
b\bar{b}\gamma\gamma$ may be also promising since the narrow
$\gamma\gamma$ peak can be reconstructed to distinguish the signal
from the backgrounds \cite{higgsself}.  Fig. \ref{cross} shows that,
for $t_{\beta}=15$ and $t_{\beta}=20$, the cross sections of
$\eta\eta$ production are only several $fb$ and even less than 1
$fb$, so it is difficult to search for $\eta\eta$ at LHC. However,
for $t_{\beta}=25$ and 35 GeV $< m_{\eta} <$ 75 GeV, the cross
sections can reach several thousand $fb$, so it is feasible to
search for $\eta\eta$ through $pp\to \eta\eta \to
\tau\bar{\tau}b\bar{b}$ at LHC.

For $m_h<150 $ GeV and $t_\beta=25$, the decay $h\to \eta\eta$ is
dominant and $h\to b\bar{b}$ is subdominant
\cite{slhv,slhhdec,slhhh}. The largest mode $pp\to \eta h\to
\eta\eta\eta$ is interesting, and a detailed study is needed to
establish the feasibility of searching for $\eta h$. For $m_h<150 $
GeV and $t_\beta=20$ or 15, the dominant decay mode is $h\to
b\bar{b}$, and $h\to \eta\eta$ is forbidden kinematically. So the
promising channel is $pp\to \eta h\to b\bar{b}\tau\bar{\tau}$. The
decay $h\to WW$ is dominant for $m_h
>150$ GeV, and the decay $h\to ZZ$  is subdominant for $m_h
>160$ GeV. Here the largest mode $b\bar{b}WW$ from the decays $\eta\to
b\bar{b}$ and $h\to WW$ has huge background $pp\to t\bar{t}\to
b\bar{b}WW$, so the mode is not optimistic \cite{bbww}. The
next-largest mode $b\bar{b}ZZ$ from the decays $\eta\to b\bar{b}$
and $h\to ZZ$ may be more promising.

Similar to the analysis of $\eta\eta$ and $\eta h$ production
processes, for $m_h<150 $ GeV and $t_\beta=25$, $pp\to
hh\to\eta\eta\eta\eta$ is the largest mode, and a detailed study is
needed to analyze the signal and the relevant backgrounds. Besides,
the promising channels are $pp\to hh\to
b\bar{b}\tau\bar{\tau}~(b\bar{b}\gamma\gamma)$ for $m_h<150 $ GeV
and $t_\beta=20$ or 15, and $pp\to hh\to WWWW$ for 150GeV $<m_h
<200$ GeV.

\section{Conclusion}
In the framework of the simplest little Higgs model, we perform a
comprehensive study for the pair productions of neutral Higgs bosons
at LHC, namely $gg(b\bar{b})\to \eta\eta$, $gg(q\bar{q})\to \eta h$
and $gg(b\bar{b})\to hh$. We find that the cross section of $gg\to
\eta\eta$ process always dominates over that of
$b\bar{b}\to\eta\eta$. The cross sections of the two processes can
reach respectively several thousand $fb$ and several hundred $fb$
when an on-shell $h$ boson mediates the two processes. When the $h$
boson is off-shell, the cross sections of the two processes are
reduced by two orders of magnitude, respectively. Besides, the cross
sections of $gg\to \eta h$ process and $q\bar{q}\to \eta h$ process
are about in the same order of magnitude, which can reach
$\ord{(10^2fb)}$ for a light $\eta$ boson. For the $hh$ production
process at LHC, the cross section in SLHM can be enhanced sizably
compared with the SM prediction. The above results imply that it is
possible to search for $\eta$ boson and $h$ boson through those
production processes at LHC. We briefly discuss the observable
signatures of $\eta\eta$, $\eta h$ and $hh$ at the LHC, and a
detailed analysis of the signatures and relevant backgrounds are
necessary in the future study.

\section*{Acknowledgment}
We thank Shuo Yang and Jin Min Yang for discussions. This work was
supported in part by the National Natural Science Foundation of
China (NNSFC) under grant No. 11005089, and by the Foundation of
Yantai University under Grant No. WL09B31.

\appendix
\section{The decay widths of h and Z'}
In addition to the SM decay modes, the $h$ boson in the SLHM can
decay into $\eta\eta$ and $Z\eta$ in the kinematically allowed
parameter space. The SLHM corrections to the tree-level decays $h\to
f \bar{f},~WW,~ZZ$ are mainly from the corresponding modified
couplings:
 \beq \Gamma(h \to XX)= \Gamma(h \to
XX)_{SM}(g_{hXX}/g_{hXX}^{SM})^2,
\end{equation}
where $XX$ denotes $WW$, $ZZ$ or fermion pairs. The SM decay width
$\Gamma(h \to XX)_{SM}$ is obtained using the code Hdecay
\cite{hdecay}. $g_{hXX}$ and $g_{hXX}^{SM}$ are the couplings of
$hXX$ in the SLHM and SM, respectively. The couplings $g_{hWW}$ and
$g_{hZZ}$ can be found in \cite{slhv}.

The decay rates of $h\to gg$ and $h\to \gamma \gamma$ are as follows
 \cite{hrrhan}:
\begin{equation}
    \begin{array}{lllll}
    \Gamma(h \to gg) & = & \displaystyle
            \frac{\alpha_s^2 m_h^3}{32 \pi^3 v^2}
            \left| \sum_f -\frac{1}{2} y_f F_{1/2}(\tau_f)
            \right|^2,
       \end{array}
\label{widthgg}
\end{equation}
where $f=t,~T,~D,~S$, $\tau_f=\frac{4m_f^2}{m_h^2}$ and
$y_f=\frac{v}{m_f}g_{hf\bar{f}}$.
\begin{equation}
    \begin{array}{lllll}
    \Gamma(h \to \gamma\gamma) & = &  \displaystyle
            \frac{\alpha^2 m_h^3}{256 \pi^3 v^2}
            \left| \sum_f y_f N_{cf} Q_f^2 F_{1/2}(\tau_{f})+\sum_V y_{_V}F_1(\tau_{_{V}})
            \right|^2,
    \end{array}
\label{widthrr}
\end{equation}
where $y_{_{V}}=\frac{v}{2m_{V}^2}g_{_{hVV}}$ with $V$ denoting the
charged gauge bosons $W^{\pm}$ and $X^{\pm}$. $N_{cf}$ and $Q_f$ are
respectively the color factor and the electric charge of the fermion
running in the loop. The dimensionless loop factors are
 \beq
    F_1(\tau) = 2 + 3 \tau + 3\tau (2-\tau) f(\tau),\quad
    F_{1/2}(\tau) = -2\tau [1 + (1-\tau)f(\tau)],
\end{equation}
where \begin{equation}
    f(\tau) = \left\{ \begin{array}{lr}
        [\sin^{-1}(1/\sqrt{\tau})]^2, & \tau \geq 1 \\
        -\frac{1}{4} [\ln(\eta_+/\eta_-) - i \pi]^2, & \, \tau < 1
        \end{array}  \right.\label{hggf12}
\end{equation}
with $\eta_{\pm}=1\pm\sqrt{1-\tau}.$

The decay rates of $h\to \eta\eta$ and $h\to Z\eta$ are \bea
\label{eq:Gamma:new}
\Gm(h \to \eta\eta) &=& \frac{{\lambda'}^2}{8\pi}\frac{v^2}{m_h} \sqrt{1-x_\eta},\nonumber\\
\Gamma( h \to Z \eta) &=& \frac{m_h^3}{32 \pi f^2}
  \left( t_\beta - \frac{1}{t_\beta} \right)^2 \,
  \lambda^{3/2} \left(1, \frac{m_Z^2}{m_h^2}, \frac{m_\eta^2}{m_h^2}
 \right ),
\eea where $x_\eta =4m_\eta^2/m_h^2$ and $\lambda (1,x,y) =
(1-x-y)^2 - 4 xy$.

The heavy gauge boson $Z'$ mainly decays into fermion pairs. \beq
\Gamma(Z'\to
f\bar{f})=\frac{N_{cf}}{24\pi}\sqrt{1-\tau_f}\left[\left((g_L^f)^2+(g_R^f)^2\right)(1-\frac{\tau_f}{4})
+\frac{3}{2}g_L^fg_R^f\tau_f\right]m_{Z'},
\end{equation}
where $g_L^f$ and $g_R^f$ are from the coupling
$Z'\bar{f}\gamma^{\mu}(g_L^f P_L + g_R^f P_R)f$, which can be found
in \cite{smoking,slhmT}.
 \beq
\Gamma(Z'\to \eta
h)=\frac{1}{96\pi}a_{Z'}^2(t_\beta-\frac{1}{t_\beta})^2\lambda^{3/2}
(1,\frac{m_\eta^2}{m_{Z'}^2},\frac{m_h^2}{m_{Z'}^2})~m_{Z'},
\end{equation}
where
$a_{Z'}=\frac{m_Z}{f}\frac{\cos\theta_W(1-\tan^2\theta_W)}{\sqrt{3-\tan^2\theta_W}}$
. Since the coupling is suppressed by $\frac{v^2}{f^2}$ and without
enhancement of $t_{\beta}$, the decay mode $Z'\to WW$ is neglected
here. The decay modes $Z'\to Y^0 Y^0$ and $Z'\to X^+ X^-$ are
forbidden kinematically \cite{etaprodecay,smoking,slhmT}.

\end{document}